\documentclass{elsart}
\usepackage{amssymb}
\usepackage{amsmath}
\usepackage[dvips]{graphicx}
\usepackage{color}

\journal{:\quad Comptes Rendus Mecanique, vol. 337 (2009)\qquad\qquad\qquad\qquad\qquad\qquad\qquad\qquad\qquad}
\input{tcilatex}
\begin{document}

\begin{frontmatter}

\title{Elastic effects of liquids on surface physics}

\author{Henri Gouin}
\ead{henri.gouin@univ-cezanne.fr}

\address{University of Aix-Marseille \& M2P2, C.N.R.S.  U.M.R.  6181, \\ Case 322, Av. Escadrille
Normandie-Niemen, 13397 Marseille Cedex 20 France}
\begin{abstract}

The contact between a liquid and an elastic solid generates a stress vector depending on the
curvature tensor in each
point of the separating surface.  For nanometer values of the mean curvature  and for suitable  materials, the  stress vector  takes significant amplitude on
the surface.  Although the surface average
action  of the liquid on the solid is the hydrostatic pressure, \emph{the
local strain generates torques} tending to regularize the surface
undulations and asperities.

\end{abstract}


\end{frontmatter}
\medskip

\section{Introduction}

\noindent In 1977, John Cahn gave simple illuminating arguments to describe
the interaction between solids and liquids. His model was based on a
generalized van der Waals theory of fluids treated as attracting hard
spheres \cite{Cahn}. It entailed assigning an energy to the solid surface
that is a functional of the liquid density \emph{at the surface}. It was
thoroughly examined in a review paper by de Gennes \cite{de Gennes}. Three
hypotheses are implicit in Cahn's picture: \ \emph{i)} \ The liquid density
is taken to be a smooth function of the distance from the solid surface,
that surface is assumed to be flat on the scale of molecular sizes and the
correlation length is assumed to be greater than intermolecular distances; \
\emph{ii)} \ The forces between solid and liquid are of short range with
respect to intermolecular distances; \ \emph{iii)} \ The liquid is
considered in the framework of a mean-field theory. This means, in
particular, that the free energy of the liquid is a classical so-called
\emph{gradient square functional}.\newline
The point of view that the liquid in an interfacial region may be treated as
bulk phase with a local free-energy density and an additional contribution
arising from the nonuniformity which may be approximated by a gradient
expansion truncated at the second order, is most likely to be successful and
perhaps even quantitatively accurate near the liquid critical point \cite%
{Rowlinson}. We use this  approximation enabling us to compute
analytically the liquid density profiles. Nevertheless, we take surface
effects and repulsive forces into account by adding density functionals at
boundary surfaces. In mean-field theory, London potentials of liquid-liquid
and liquid-solid molecular interactions are
\begin{equation*}
\left\{
\begin{array}{c}
\displaystyle\;\;\;\;\;\;\varphi _{ll}=-\frac{c_{ll}}{r^{6}}\;,\text{ \
when\ }r>\sigma _{l}\;\;\text{and }\;\ \varphi _{ll}=\infty \text{ \ when \ }%
r\leq \sigma _{l}\, ,\  \\
\displaystyle\;\;\;\;\;\;\varphi _{ls}=-\frac{c_{ls}}{r^{6}}\;,\text{ \
when\ }r>\delta \;\;\text{and }\;\ \varphi _{ls}=\infty \text{ \ when \ }%
r\leq \delta \;, \
\end{array}
\right.
\end{equation*}
where $c_{ll}$ et $c_{ls}$ are two positive constants associated with
Hamaker constants, $\sigma _{l}$ and $\sigma _{s}$ respectively denote
liquid and solid molecular diameters, $\delta =\frac{1}{2}($ $\sigma _{l}+$ $%
\sigma _{s})$ is the minimal distance between centers of liquid and solid
molecules \cite{Israelachvili}.

We consider the interaction between a solid surface flat at a molecular
scale (but curved at several nanometer scale) and a liquid by
means of a continuous model. The density-functional of energy $E$ of the
inhomogeneous liquid in a domain $D$ of differentiable boundary $S$
(external forces being neglected) is taken in the form
\begin{equation*}
E=E_{f}+E_{S}\qquad {\mathrm{with}\qquad \emph{E}_{\emph{f}}=\int \int
\int_{D}\rho \,\varepsilon  \ dv ,\ \emph{E}_{\emph{S}}\ =\
\int \int_{S}\phi \ ds} .\label{E}
\end{equation*}%
The first integral (energy of the volume) is associated with square-gradient
approximation when we introduce a specific free energy of the fluid at a
given temperature, $\varepsilon =\varepsilon (\rho ,\beta )$, as a function
of liquid density $\rho $ and $\beta =(\mathrm{grad\,\rho )^{2}}$. Specific
free energy $\varepsilon $ characterizes together fluid properties of \emph{%
compressibility} and \emph{molecular capillarity} of interfaces. In
accordance with gas kinetic theory \cite{Rocard}, scalar $\lambda =2\rho
\,\varepsilon _{,\beta }(\rho ,\beta )$ (where $\varepsilon _{,\beta }$
denotes the partial derivative with respect to $\beta $) is assumed to be
constant at a given temperature and
\begin{equation*}
\rho \,\varepsilon =\rho \,\alpha (\rho )+\frac{\lambda }{2}\,(\text{grad\ }%
\rho )^{2},
\end{equation*}%
where term $({\lambda }/{2})\,(\mathrm{grad\ \rho )^{2}}$ is added to the
volume free energy $\rho \,\alpha (\rho )$ of a compressible fluid. We
denote the pressure term by $P(\rho )=\rho ^{2}\alpha ^{\,\prime }(\rho ). $
The second integral (energy of the surface) is such that the free energy per
unit surface $\phi $ is \cite{de Gennes},
\begin{equation}
\phi (\rho )=-\gamma _{1}\rho +\frac{1}{2}\,\gamma _{2}\,\rho ^{2}.
\label{surface energy}
\end{equation}%
Here $\rho $ denotes the limit liquid density value at surface $S$.
Constants $\gamma _{1}$, $\gamma _{2}$ and $\lambda $ are positive and given
by relations \cite{Gouin},
\begin{equation*}
\gamma _{1}=\frac{\pi c_{ls}}{12\delta ^{2}m_{l}m_{s}}\;\rho _{sol},\quad
\gamma _{2}=\frac{\pi c_{ll}}{12\delta ^{2}m_{l}^{2}},\quad \lambda =\frac{%
2\pi c_{ll}}{3\sigma _{l}\,m_{l}^{2}},  \label{coefficients}
\end{equation*}%
where $m_{l}$ et $m_{s}$ respectively denote the masses of liquid and solid
molecules; $\rho _{sol}$ is the solid density.

In this paper, we first develop the boundary conditions for the general case
of the interaction between a non-homogeneous liquid and a curved solid
surface with a surface energy due to intermolecular interactions and
depending of the fluid volume deformation. Then, for a surface energy in
form (\ref{surface energy}) we study the stress vector distribution on a
surface where bumps and hollows are periodically distributed. Finally, we
estimate the stress effects for a silicon surface, with a curvature of
several nanometer range, in contact with water.

\section{Boundary conditions}

The equation of equilibrium and boundary conditions are obtained by using
the virtual power principle \cite{Germain,Maugin}. For example, virtual
displacements ${\mathbf{\zeta}} =\delta \mathbf{x}$ are defined in a
classical way by Serrin \cite{Serrin} page 145, where $\mathbf{x}%
=\{x^{i}\},(i=1,2,3)$ denotes the Euler variables in a Galilean or fixed
system of coordinates.\newline
\begin{figure}[h]
\begin{center}
\includegraphics[width=7cm]{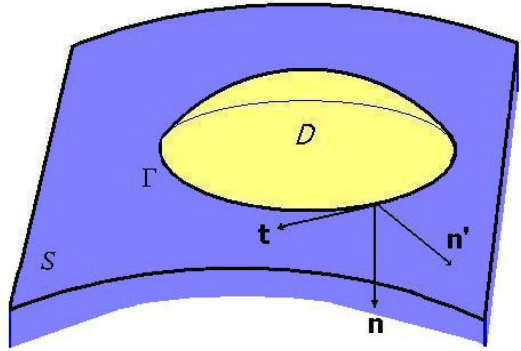}
\end{center}
\caption{ Vector $\mathbf{n}$ is the unit normal vector to $S$ exterior to $%
D $; vector $\mathbf{t}$ is the unit tangent vector to $\Gamma $ with
respect to $\mathbf{n}$; ${\mathbf{n}^{\prime }}={\mathbf{t}}\times {\mathbf{%
n}}$.}
\label{fig1}
\end{figure}
A liquid (in drop form) occupying a domain $D$ of the physical space lies on
a solid surface $S$ (the liquid is also partially bordered by a gas); the
edge $\Gamma $ (or contact line) is the curve common to $S$ and the boundary
of $D$ (see Fig. 1). All the surfaces and curves are oriented differential
manifolds (\footnote{\textit{Transposed} mappings being denoted by $^{T}$,
for any vectors ${\mathbf{a}},{\mathbf{b}}$, we write ${\mathbf{a}}^{T}\,%
\mathbf{b}$\ for their \textit{scalar product} (the line vector is
multiplied by the column vector) and ${\mathbf{a}}{\mathbf{b}}^{T}$ or ${%
\mathbf{a}}\otimes {\mathbf{b}}$ for their \textit{tensor product} (the
column vector is multiplied by the line vector). The image of vector ${%
\mathbf{a}}$ by a mapping $B$ is denoted by $B\,{\mathbf{a}}$. Notation ${%
\mathbf{b}}^{T}\,{B}\,$ means the covector ${\mathbf{c}}^{T}$ defined by the
rule ${\mathbf{c}}^{T}=(B^{T}\,{\mathbf{b}})^{T}$. The divergence of a
linear transformation $B $ is the covector $\mathrm{div}B$ such that, for
any constant vector ${\mathbf{a}},$ $(\mathrm{div}\,B)\,{\mathbf{a}}=\mathrm{%
div}\,(B\ {\mathbf{a}})$. If $f$ is a real function of $\mathbf{x}$, $%
\displaystyle{\partial f}/{\partial {\mathbf{x}}}$ is the linear form
associated with the gradient of $f $ and $\displaystyle{\partial f}/{%
\partial x^{i}}=({\partial f}/{\partial {\mathbf{x}}})_{i}$\thinspace ;
consequently, $\displaystyle({\partial f}/{\partial {\mathbf{x}}})^{T}=%
\mathrm{grad}\,f$. The identity tensor is denoted by $I$.}).

\subsection{Variation of the density-functional of energy $E$}
The density in the fluid has a limit value at the wall $S$. Then, on $S$,
\begin{equation*}
\delta \phi =\phi ^{\prime }(\rho )\,\delta \rho =-\rho \,\phi ^{\prime
}(\rho )\,\mathrm{div}\,{\mathbf{\zeta }},
\end{equation*}%
where $\delta \rho +\rho \,\mathrm{div}\,{\mathbf{\zeta }}=0$ \cite{Serrin}.
Let us denote
\begin{equation*}
G=-\rho \,\phi ^{\prime }(\rho )\,,\quad H=\phi (\rho )-\rho \,\phi ^{\prime
}(\rho ).
\end{equation*}%
The function $H$ is the Legendre transform of $\phi$ with respect to $\rho$.
For any virtual displacement ${\mathbf{\zeta }}$ null on $\Gamma $, Rel. (\ref%
{A1}) of Appendix yields,
\begin{equation*}
\int \int_{S}\delta \phi \ ds=\int \int_{S}G\ \mathrm{div}\,{\mathbf{\zeta }}%
\ ds\equiv \displaystyle\int \int_{S}\ \left\{ G\,\frac{d\zeta _{n}}{dn}%
-\left( \frac{2G}{R_{m}}\,{\mathbf{n}}^{T}+\mathrm{grad}_{tg}^{T}G\right) {%
\mathbf{\zeta }}\right\} ds ,
\end{equation*}
where, now $S$ is the imprint of $D$ on the solid surface.
Consequently, from the calculations in Appendix, we obtain:\newline
\emph{For any virtual displacement null on the complementary boundary of $D$
with respect to $S$ and null on the edge $\Gamma $}, the variation of $E$
is,
\begin{equation*}
\begin{array}{ll}
\delta E=\displaystyle-\int \int \int_{D}(\mathrm{div}\,\sigma )\,{\mathbf{%
\zeta }}\,dv\  &  \\
+\displaystyle\int \int_{S}(G-A)\frac{d{\zeta }_{n}}{dn}+\left( \frac{2(A-H)%
}{R_{m}}\,{\mathbf{n}}^{T}+\mathrm{grad}_{tg}^{T}(A-H)+{\mathbf{n}}%
^{T}\sigma \right) {\mathbf{\zeta }}\ ds, \label{S3} &
\end{array}%
\end{equation*}
where
\begin{equation}
\displaystyle\ \sigma =-p\,{I}-\lambda \ \mathrm{grad}\;\rho \otimes \mathrm{%
grad}\;\rho \equiv -p\,{I}-\lambda \,\left( \frac{\partial \rho }{\partial {%
\mathbf{x}}}\right) ^{T} \frac{\partial \rho }{\partial {\mathbf{x}}}
\label{sigma}
\end{equation}
is the symmetric stress tensor of the inhomogeneous liquid, with $p=\rho
^{2}\varepsilon _{,\rho }-\rho \ \mathrm{div}(\lambda \mathrm{\ grad}\,\rho
) $ ; $A=\lambda \,\rho \,\left({d\rho }/{dn}\right) $ with$\ {d\rho }/{dn}
=\left({\partial\rho }/{\partial{\mathbf{x}}}\right)\, {\mathbf{n}}$ ; ${%
\zeta }_{n}={\mathbf{n}}^{T}\,{\mathbf{\zeta }}$ ; $2/R_m$ is the mean
curvature of $S$ and $\mathrm{grad}_{tg}$ denotes the tangential part of the
gradient relatively to $S$.
\subsection{The virtual work of forces exerted on $D$}
The virtual work of elastic stresses on $S$ is
\begin{equation*}
\delta \tau _{e}=\int \int_{S}{\mathbf{\kappa }}^{T}{\mathbf{\zeta }}\ ds\,,
\end{equation*}%
where ${\mathbf{\kappa }}={-\sigma _{e}}\,{\mathbf{n}}$ is the loading
vector associated with stress tensor $\sigma _{e}$ on the wall in classical
theory of continuum mechanics (\footnote{%
It is important to note that the external unit normal to $S$ with respect to
the solid is $\mathbf{-n}$.}). Then, the virtual work of forces $\delta \tau
$ exerted on $D$ is
$
\delta \tau =-\,\delta E+\delta \tau _{e}\,
$ and,
\begin{equation*}
\begin{array}{ll}
\delta \tau=\displaystyle\int \int \int_{D}(\mathrm{div}\,\sigma )\,{\mathbf{%
\zeta }}\,dv\  &  \\
-\displaystyle\int \int_{S}(G-A)\frac{d{\zeta }_{n}}{dn}+\left( \frac{2(A-H)%
}{R_{m}}\,{\mathbf{n}}^{T}+\mathrm{grad}_{tg}^{T}(A-H)+{\mathbf{n}}%
^{T}\sigma-{\mathbf{\kappa }}^{T} \right) {\mathbf{\zeta }}\ ds.\label{S3} &
\end{array}%
\end{equation*}

\subsection{Results}

The fundamental lemma of variation calculus,
applied to the relation $\delta \tau =0$, for all previous virtual displacements, yields:

$\bullet$ \ The well-known equation of equilibrium for
capillary fluids \cite{Casal},
\begin{equation}
\mathrm{div}\,\sigma =0,  \label{motion5}
\end{equation}%
$\bullet$ \ The boundary conditions on $S$,
\begin{equation}
\forall \ {\mathbf{x}}\in S,\ \ \ \ \ \ \ \ \ \left\{
\begin{array}{l}
G-A=0, \\
{\mathbf{\kappa }}=\,\displaystyle\frac{2(A-H)}{R_{m}}\;{\mathbf{n}}+\mathrm{%
grad}_{tg}(A-H)+\sigma \;{\mathbf{n}}.%
\end{array}%
\right.  \label{conditions}
\end{equation}
Equation (\ref{conditions})$_1$ yields a condition relative to
the surface energy (\ref{surface energy}) which depends on the fluid density
at the surface and on the quality of the solid wall:
\begin{equation}
\lambda \ \frac{d\rho }{dn}+\phi ^{\prime }(\rho )=0 \qquad \rm{or}\qquad  \lambda \,\frac{d\rho }{dn}=\gamma _{1}-\gamma _{2}\,\rho \,. \label{conditions5.1}
\end{equation}%
Equation (\ref{conditions5.1}) expresses an embedding effect for the liquid
density. Such a condition appears for simpler geometry in \cite%
{Cahn,Seppecher}. \newline
Condition (\ref{conditions})$_2$ appears in the literature \cite%
{Germain,Casal2} but without the  terms  corresponding to the molecular
model (\ref{surface energy}) of surface free energy. Such type of condition
also appears in interfacial problems with other solid surface energy but
with a null curvature as in \cite{Cahn,Seppecher}. In Cauchy theory, we are
back to the classical equation ${\mathbf{\kappa }}=\sigma \;{\mathbf{n}}$.%
\newline
The definition (\ref{sigma}) of $\sigma $ implies\
$
\sigma \,{\mathbf{n}}=-p\,{\mathbf{n}}-\lambda \; ({d\rho }/{dn})\ \mathrm{%
grad}\;\rho \,.
$
Then, for an elastic wall, by taking into account of Rel.  (\ref{conditions5.1}), the vector $\mathbf{\kappa }$ is normal to $S$,
\begin{equation}{\mathbf{\kappa }}=\kappa _{n}\,{\mathbf{n}} \quad \rm{with}\quad
\kappa _{n}\ =\mathbf{n}^T\sigma\,\mathbf{n}-\frac{2\,\phi }{R_{m}}\, .  \label{conditions5.2}
\end{equation}%
We obtain the stress vector values of the solid at the elastic wall (which
is opposite to the action ${\mathbf{\tau }}$ of the liquid on the elastic
wall). Relation (\ref{conditions5.2})  looks like  the Laplace
formula for fluid interfaces. Nonetheless, we will see in the
next section some differences between the results for fluid interfaces and for liquid-solid interfaces.

\section{An example of elastic effect on a solid surface}

\subsection{General considerations}

As bibliography about elastic effects on surface physics, one may refer to
the review article \cite{Muller}.\newline
The aim of this section is to present an example of system such that the
mesoscopic effects of a liquid locally generate important molecular stress
vectors on a solid surface. We consider a periodic domain such that the
substrate solid surface has an alternated structure. The solid surface can
be considered as a flat domain at the Angstr\"{o}m scale because roughness
and undulations are only of several nanometer length (such a model is
presented on Fig. 2).
\begin{figure}[h]
\begin{center}
\includegraphics[width=8cm]{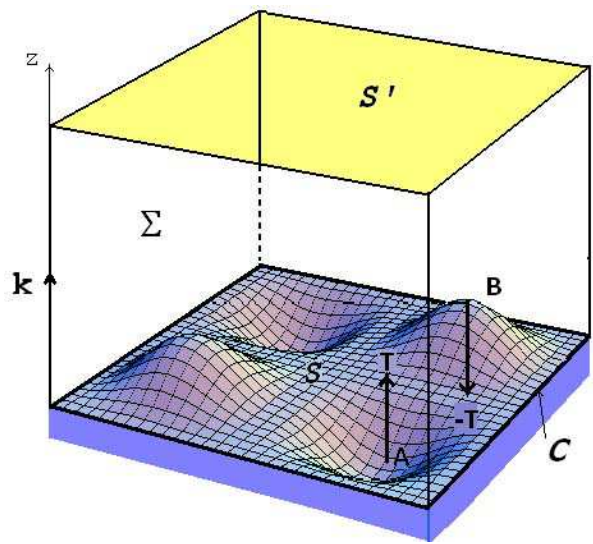}
\end{center}
\caption{ We consider the model consisting of a surface $S$ with bumps and
hollows periodically distributed on a period $L$ of several nanometers in
two directions such that, with respect to the third axis, the bump and hollow
levels are opposite. Extrema of the surface mean curvature are located at
point A and B; curve $C$ is the limit curve of the periodic rectangular
parallepiped. Surface $S^{\prime }$ delimits the
liquid bulk (at a distance $h$ of a great number of nanometers from surface $%
S$). Surface $\Sigma $ is the lateral boundary of $D$. Vector $\mathbf{k}$
is normal to $S^{\prime }$ and  $z$ is directed along $\mathbf{k}$.}
\label{fig2}
\end{figure}
At level 0 with respect to the third axis, the lateral boundary of domain $D$
follows the curve $C$ of the bludging surface. Due to the axial symmetries
around the lines A$\mathbf{k}$ and B$\mathbf{k}$, in local coordinates with
these lines as third axis, $\mathrm{{grad}\,\rho =(d\rho /dz)\mathbf{k}}$
and on these lines the stress tensor $\sigma $ of the inhomogeneous liquid
gets expressions in the form
\begin{equation*}
\mathbf{\sigma }=\left[
\begin{array}{ccc}
a_{1} & 0 & 0 \\
0 & a_{2} & 0 \\
0 & 0 & a_{3}%
\end{array}%
\right] ,\quad \mathrm{with}\quad \left\{
\begin{array}{lll}
a_{1} & = & a_{2}=-p,\quad \displaystyle p=P(\rho )-\frac{\lambda }{2}\left(
\frac{d\rho }{dz}\right) ^{2}-\lambda \,\rho \,\Delta \rho \\
a_{3} & = & \displaystyle{-p-\lambda \left( \frac{d\rho }{dz}\right) ^{2},}%
\end{array}%
\right.
\end{equation*}
where $\Delta$ is the Laplace operator.
Consequently, on these lines, Eq. (\ref{motion5}) yields a constant value
for the eigenvalue $a_{3}$,
\begin{equation}
p+\lambda \left( \frac{d\rho }{dz}\right) ^{2}=P_{l},  \label{equilibrium1b}
\end{equation}%
where $P_{l}$ denotes the uniform pressure in the liquid bulk of density $%
\rho _{l}$ bounding the liquid layer at level $h$. \newline
-\ \ Due to symmetries of domain $D$, we deduce the average stress
actions of the liquid on $S$ and $S^{\prime }$ are opposite and numerically
equal to the pressure $P_{l}$. \newline
-\ \ From Rels. (\ref{conditions5.1}-\ref{equilibrium1b}) we obtain, at
points A and B, a stress vector ${\mathbf{\tau }}=-{\mathbf{\kappa }}$,
action of the liquid on the elastic wall in the same form than the Laplace
formula form for fluid interfaces,
\begin{equation}
{\mathbf{\tau }}=\left( P_{\ell }+\frac{2\,\phi }{R_{m}}\right) \,\mathbf{n}.
\label{Laplace}
\end{equation}%
-\ \ We must emphasize that Rel. (\ref{Laplace}) is only valid at points A
and B. In fact,   Rel. (\ref{conditions5.2})  yields
\begin{equation}
{\mathbf{\tau }}=\left( -\mathbf{n}^T\sigma\,\mathbf{n} +\frac{2\,\phi }{R_{m}}\right)\,\mathbf{n}, \label{Laplacebis}
\end{equation}%
but for points which are not the summits of bumps or the bottoms of hollows,
 $-\mathbf{n}^T\sigma\,\mathbf{n}\equiv p+\lambda \left({d\rho }/{dn}\right) ^{2} \neq P_{\ell }$\ where\ $\lambda({d\rho }/{dn})^2 \neq \lambda\left( \mathrm{%
grad}\rho\right) ^{2}$. Consequently at a
mesoscopic scale, due to the anisotropy of the liquid on curved solid
surfaces, Rel. (\ref{Laplacebis}) replaces Laplace's formula of fluid
interfaces. \newline
-\ \ The stress vector is directed as $\mathbf{k}$ at points A and B. Due to
the axial symmetries around the surface extrema at points A and B and
opposite mean curvatures, when we neglect $P_{l}$ with respect to $2\phi/R_m$, the stress vector associated with the hollow
corresponding to point A is a vector $\mathbf{T}$ parallel to $\mathbf{k}$
and the stress vector associated with the bump corresponding to point B is a
vector $\mathbf{-T}$; the two vectors generate a torque on the surface. This
result is in accordance with results in \cite{Prevot} where the interaction between liquid and solid is represented as
localized dipoles and monopoles depending on bumps and hollows of the
surface $S$.

\subsection{Application to explicit materials}

At $\theta= 20 {{}^\circ}$ Celsius, we consider water damping a wall in
silicon. The experimental estimates of coefficients defined in Section 1 are
presented in Table 1. \newline
Far from the liquid critical point, the liquid density at the wall is
closely the same than the liquid density in the bulk \cite{Gouin1}.

If we
consider a mean radius of curvature of surface S, $R_m=-10^{-6}$ cm at point
A and $R_m=10^{-6}$ cm at point B, when we neglect $P_l$, we immediately obtain an arithmetic value
of $\tau_{n} \equiv {\mathbf{n}}^T{\mathbf{\tau }}= 10^8$ cgs (or 100 atmospheres) corresponding to stress
effects of large magnitude between areas around points A and B.

\begin{table}[tbp]
\centering
$%
\begin{tabular}{|c|c|c|c|c|}
\hline\hline
\multicolumn{1}{||c|}{\scriptsize Physical constants} & $c_{ll}$ & $%
\sigma_{l}$ & $m_{l}$ & \multicolumn{1}{|c||}{$\rho_{l}$} \\ \hline
\multicolumn{1}{||c|}{Water} & $1.4\times 10^{-58}$ & $2.8\times 10^{-8}$ & $%
2.99\times 10^{-23}$ & \multicolumn{1}{|c||}{$0.998$} \\ \hline\hline
\multicolumn{1}{||c|}{\scriptsize Physical constants} & $c_{ls}$ & $%
\sigma_{s}$ & $m_{s}$ & \multicolumn{1}{|c||}{$\rho_{sol}$} \\ \hline
\multicolumn{1}{||c|}{Silicon} & $1.4\times 10^{-58}$ & $2.7\times 10^{-8}$
& $4.65\times 10^{-23}$ & \multicolumn{1}{|c||}{$2.33$} \\
\hline\hline\hline\hline
\multicolumn{1}{||c|}{\scriptsize Deduced constants} & $\delta$ & $\lambda$
& $\gamma_1$ & \multicolumn{1}{|c||}{$\gamma_2$} \\ \hline
\multicolumn{1}{||c|}{\scriptsize Results (water-silicon)} & $2.75\times
10^{-8}$ & $1.17\times 10^{-5}$ & $81.2$ & \multicolumn{1}{|c||}{$54.2$} \\
\hline\hline
\end{tabular}
$ \vskip 0.5cm
\caption{ The physical values associated with water and silicon are obtained
in references \protect\cite{Israelachvili,Handbook} and expressed in \textbf{%
c.g.s. units} (centimeter, gramme, second). No information is available for
water-silicon interactions; we assume that $c_{ll}=c_{ls}$.}
\label{TableKey1}
\end{table}
The elastic effects of a liquid on a solid surface result from the topology
of the contact interface. It is amazing to observe that a solid surface
considered as an interface between solid and liquid does not require new
concept but only a supplementary surface energy and likewise surface
morphology.\newline
An important assumption in the previous calculations is that three scales
infer in the surface physics: a length scale  of one nanometer associated
with molecular effects and the expression of surface energy, a length scale
of ten nanometers associated with the size of undulations and surface
roughness and a  length scale of one hundred nanometers associated with the
distance of the liquid bulk to the surface $S$.

\section{Appendix}

Let $S$ be a differentiable oriented manifold in the 3-dimensional space and
${\mathbf{n}}$ its oriented unit normal locally extended in the vicinity of $%
S$ by the expression ${\mathbf{n(x)}}=\mathrm{grad}\ d(\mathbf{x}),$ where $%
d(\mathbf{x}) $ is the distance of point $\mathbf{x}$ to $S$; covectors $%
\mathrm{grad}^{T}a $ and $\mathrm{grad}_{tg}^{T}a$ denote the transposition
of $\mathrm{grad}\,a$ and $\mathrm{grad}_{tg}a$, respectively; for any
vector field ${\mathbf{w}}$, we get \cite{Germain}:\newline
{$\displaystyle\mathrm{rot}({\mathbf{n}}\times {\mathbf{w}})={\mathbf{n}}\,%
\mathrm{div}\,{\mathbf{w}}-{\mathbf{w}}\,\mathrm{div}\,{\mathbf{n}}+\frac{%
\partial {\mathbf{n}}}{\partial {\mathbf{x}}}\,{\mathbf{w}}-\frac{\partial {%
\mathbf{w}}}{\partial {\mathbf{x}}}\,{\mathbf{n}}.$}\newline
From $\,\displaystyle{\mathbf{n}}^{T}\frac{\partial {\mathbf{n}}}{\partial {%
\mathbf{x}}}=0\,$ and $\,\mathrm{div}\,{\mathbf{n}}=\displaystyle-\frac{2}{%
R_{m}}$ we obtain on $S$:\newline
\begin{equation}
{\mathbf{n}^{T}}\mathrm{rot}({\mathbf{n}}\times {\mathbf{w}})=\mathrm{div}\,{%
\mathbf{w}}+\frac{2}{R_{m}}\,{\mathbf{n}^{T}}{\mathbf{w}}-{\mathbf{n}^{T}}%
\frac{\partial {\mathbf{w}}}{\partial {\mathbf{x}}}\,{\mathbf{n}},\newline
\label{A0}
\end{equation}%
and we deduce,

\begin{lem}
For any differentiable scalar field $a,$%
\begin{equation}
a\,\displaystyle\mathrm{div}\,{\mathbf{\zeta }}=a\,\frac{d{\zeta }_{n}}{dn}-%
\frac{2a}{R_{m}}\,{\zeta }_{n}-(\mathrm{grad}_{tg}^{T}a)\,{\mathbf{\zeta }}+{%
\mathbf{n}}^{T}\mathrm{rot}\,(a{\mathbf{n}}\times {\mathbf{\zeta }}).
\label{A1}
\end{equation}%
where $\displaystyle\mathrm{grad}_{tg}^{T}a=\left[ \frac{\partial a}{%
\partial \mathbf{x}}\left( {I-\mathbf{nn}}^{T}\right) \right] $ belongs to
the cotangent plane to $S$ and $\displaystyle\frac{d{\zeta }_{n}}{dn}={%
\mathbf{n}}^{T}\frac{\partial {\mathbf{\zeta }}}{\partial \mathbf{x}}{%
\mathbf{n.}}$
\end{lem}

$\bullet$\quad {\textbf{Application to the calculation of} $\delta E_{f}$: }

\emph{All the densities are expressed in the physical space.} The domain $D$
is a material volume \cite{Serrin}, then $\ \displaystyle\delta E_{f}=\int
\int \int_{D}\rho \ \delta \varepsilon \,dv$.\newline
From $\displaystyle\ \delta \varepsilon =\frac{\partial \varepsilon }{%
\partial \rho }\ \delta \rho +\frac{\partial \varepsilon }{\partial \beta }\
\delta \beta $ and $\displaystyle\delta \,\frac{\partial \rho }{\partial {%
\mathbf{x}}}=\frac{\partial \delta \rho }{\partial {\mathbf{x}}}-\frac{%
\partial \rho }{\partial {\mathbf{x}}}\,\frac{\partial {\mathbf{\zeta }}}{%
\partial {\mathbf{x}}}$\ \ (see\ \ \cite{Casal}), we get:
\begin{eqnarray*}
\rho \,\varepsilon _{,\beta }\,\delta \beta &=&2\rho \,\varepsilon _{,\beta
}\,\delta \frac{\partial \rho }{\partial {\mathbf{x}}}\,\left( \frac{%
\partial \rho }{\partial {\mathbf{x}}}\right) ^{T}\equiv \lambda \left(
\frac{\partial \delta \rho }{\partial {\mathbf{x}}}-\frac{\partial \rho }{%
\partial {\mathbf{x}}}\,\frac{\partial {\mathbf{\zeta }}}{\partial {\mathbf{x%
}}}\right) \left( \frac{\partial \rho }{\partial {\mathbf{x}}}\right) ^{T} \\
&\equiv &\mathrm{div}(\lambda \ \mathrm{grad}\,\rho \ \delta \rho )-\mathrm{%
div}(\lambda \ \mathrm{grad}\,\rho )\,\delta \rho -tr\left( \lambda \,%
\mathrm{grad}\,\rho \ \mathrm{grad}^{T}\rho \ \frac{\partial {\mathbf{\zeta }%
}}{\partial {\mathbf{x}}}\right) .
\end{eqnarray*}%
Due to $\delta \,\rho =-\rho \,\mathrm{div}\,{\mathbf{\zeta }}$\ \ (see\ \
\cite{Serrin}), with $a=\lambda \,\rho \,(d\rho /dn)\equiv A$ and by using
definition (\ref{sigma}), we obtain:
\begin{equation*}
\begin{array}{ll}
\displaystyle\delta E_{f}=\int \int \int_{D}\left( \frac{\partial p}{%
\partial {\mathbf{x}}}+\mathrm{div}(\lambda \ \mathrm{grad}\ \rho \ \mathrm{%
grad}^{T}\rho )\right) {\mathbf{\zeta }}\,dv &  \\
\displaystyle\qquad -\int \int \int_{D}\mathrm{div}\left( \lambda \,\rho \,%
\mathrm{grad}\,\rho \ \mathrm{div}\,{\mathbf{\zeta }}+\lambda \,\mathrm{grad}%
\,\rho \ \mathrm{grad}^{T}\rho \ {\mathbf{\zeta }}+p\,{\mathbf{\zeta }}%
\right) dv &  \\
\displaystyle\qquad\equiv \int \int \int_{D}-(\mathrm{div}\,\sigma )\,{%
\mathbf{\zeta }}\,dv+\int \int_{S}(-A\ \mathrm{div}\,{\mathbf{\zeta }}+{%
\mathbf{n}}^{T}\sigma \,{\mathbf{\zeta }})\,ds. &
\end{array}%
\end{equation*}%
From the Stokes formula, we get:%
\begin{equation*}
\int \int_{S}{\mathbf{n}}^{T}\mathrm{rot}\,(A{\mathbf{n}}\times {\mathbf{%
\zeta }})\,ds = \int_{\Gamma }A\,{\mathbf{t}}^{T}\,\left( {\mathbf{%
n\times \zeta }}\right) \,d\ell \equiv \int_{\Gamma }A\,{\mathbf{n}^{\prime }%
}^{T}\,{\mathbf{\zeta }}\,d\ell,
\end{equation*}
which is null in the case of the virtual displacements  of Section 2.1. Finally, by using Rel. (\ref{A1}),
\begin{equation*}
\delta E_{f}=\int \int \int_{D}-(\mathrm{div}\,\sigma )\,{\mathbf{\zeta }}%
\,dv+\int \int_{S}\left\{ -A\,\frac{d\zeta _{n}}{dn}+\left( \frac{2A}{R_{m}}%
\,{\mathbf{n}}^{T}+\mathrm{grad}_{tg}^{T}\,A+{\mathbf{n}}^{T}\sigma \right) {%
\mathbf{\zeta }}\right\} ds.
\end{equation*}%
$\bullet $\quad {\textbf{Application to the calculation of}} $\delta E_{S}$:

Due to $\displaystyle\ E_{S}=\int \int_{S}\phi \ \det \,({\mathbf{n}},d_{1}{%
\mathbf{x}},d_{2}{\mathbf{x}})\ $ where $\displaystyle\ d_{1}{\mathbf{x}}\ $
and $\displaystyle\ d_{2}{\mathbf{x}}\ $ are two coordinate lines of $S,$ we
get:
\begin{equation*}
E_{S}=\int \int_{S_{0}}\phi \,\det \,F\ \hbox{det}\,(F^{-1}{\mathbf{n}},d_{1}%
{\mathbf{X}},d_{2}{\mathbf{X}}),
\end{equation*}%
where $S_{0}$ is the image of $S$ in a reference space with Lagrangian
coordinates ${\mathbf{X}}$ and $F$ is the deformation gradient tensor $%
\displaystyle\frac{\partial {\mathbf{x}}}{\partial {\mathbf{X}}}$ of
components $\left\{ \displaystyle\frac{\partial {x^{i}}}{\partial {X^{j}}}%
\right\} $, (see \cite{Serrin}). Then,
\begin{equation*}
\displaystyle\delta E_{S}=\int \int_{S_{0}}\delta \phi \,\det F\,\hbox{det}%
(F^{-1}{\mathbf{n}},d_{1}{\mathbf{X}},d_{2}{\mathbf{X}})+\int
\int_{S_{0}}\phi \ \delta \left( \det F\;\hbox{det}(F^{-1}{\mathbf{n}},d_{1}{%
\mathbf{X}},d_{2}{\mathbf{X}})\right) ,
\end{equation*}%
\begin{equation*}
\begin{array}{ll}
\mathrm{with}\ \ \displaystyle\int \int_{S_{0}}\phi \ \delta \left( \det
\,F\ \hbox{det}\,(F^{-1}{\mathbf{n}},d_{1}{\mathbf{X}},d_{2}{\mathbf{X}}%
)\right)  &  \\
\quad \ =\displaystyle\int \int_{S}\phi \ \mathrm{div}\,{\mathbf{\zeta }}\
\det ({\mathbf{n}},d_{1}{\mathbf{x}},d_{2}{\mathbf{x}})+\phi \,\det \left(
\frac{\partial {\mathbf{n}}}{\partial {\mathbf{x}}}\,{\mathbf{\zeta }},d_{1}{%
\mathbf{x}},d_{2}{\mathbf{x}}\right) -\phi \,\det \left( \frac{\partial {%
\mathbf{\zeta }}}{\partial {\mathbf{x}}}\,{\mathbf{n}},d_{1}{\mathbf{x}}%
,d_{2}{\mathbf{x}}\right)  &  \\
\quad \ =\displaystyle\int \int_{S}\left( \mathrm{div}(\phi \,{\mathbf{\zeta
}})-(\mathrm{grad}^{T}\phi )\,{\mathbf{\zeta }}-\phi\, {\mathbf{n}}^{T}\frac{%
\partial {\mathbf{\zeta }}}{\partial {\mathbf{x}}}\,{\mathbf{n}}\right) ds.
&
\end{array}%
\end{equation*}%
Relation (\ref{A0}) yields:
\begin{equation*}
\displaystyle\mathrm{div}\,(\phi \,{\mathbf{\zeta }})+\frac{2\,\phi }{R_{m}}%
\,{\mathbf{n}}^{T}{\mathbf{\zeta }}-{\mathbf{n}}^{T}\frac{\partial \phi \,{%
\mathbf{\zeta }}}{\partial {\mathbf{x}}}\,{\mathbf{n}}={\mathbf{n}}^{T}\,%
\mathrm{rot}\,(\phi \,{\mathbf{n}}\times {\mathbf{\zeta }}).
\end{equation*}%
Consequently,
\begin{equation*}
\begin{array}{ll}
\displaystyle\quad \ \int \int_{S_{0}}\phi \,\delta \left( \det \ F\,%
\hbox{det}\,(F^{-1}{\mathbf{n}},d_{1}{\mathbf{X}},d_{2}{\mathbf{X}})\right)
&  \\
\displaystyle\ =\int \int_{S_{0}}\left( -\frac{2\,\phi }{R_{m}}\,{\mathbf{n}}%
^{T}+\mathrm{grad}^{T}\phi \,({\mathbf{nn}^{T}-I})\right) {\mathbf{\zeta }}%
\,ds+\int \int_{S}{\mathbf{n}}^{T}\ \mathrm{rot}\,(\phi \,{\mathbf{n}}\times
{\mathbf{\zeta }})\,ds, &
\end{array}%
\end{equation*}%
and finally due to $\displaystyle\int_{\Gamma }\phi \,{\mathbf{n}^{\prime }}%
^{T}{\mathbf{\zeta }}\,d\ell =0$ for the virtual displacements of Section 2.1,
\begin{equation*}
\delta E_{S}=\int \int_{S}\left( \delta \phi -\left( \frac{2\,\phi }{R_{m}}\
{\mathbf{n}}^{T}+\mathrm{grad}_{tg}^{T}\phi \right) {\mathbf{\zeta }}\right)
ds.
\end{equation*}

\end{document}